Appears in *Digital Signal Processing* https://doi.org/10.1016/j.dsp.2020.102743 (vol. 102, July 2020, 102743)
This version (with a typo in Eq. 4b corrected) is available online at arXiv:1912.08341 [eess.SY]

`

# Improving the frequency response of Savitzky-Golay filters via colored-noise models

Hugh L. Kennedy

DST Group, ISS Division, Edinburgh, SA 5111 Australia
Hugh.Kennedy@DST.defence.gov.au

*Abstract*— **Savitzky-Golay (SG) filters are finite impulse response (FIR) realizations of least-squares polynomial regression and they are widely used for filtering (e.g. smoothing, interpolating, predicting, differentiating) and processing (e.g. detecting and classifying) non-stationary signals in non-Gaussian noise. For such inputs, the Wiener filter is biased and the Kalman filter is sub-optimal. Sequentially-correlated (i.e. 'colored') noise models are an integral part of the Wiener filter and an optional addition to the Kalman filter; however, their use in SG-filters has been overlooked in recent times. It is shown here that colored (wide-band and narrow-band) noise models are readily incorporated into a standard SG-filter and that this also addresses the well-known deficiency of their poor frequency-selectivity/configurability. A wide-band noise model sets the main-lobe/side-lobe width/height and provides physical justification for band-limited design procedures described elsewhere. The proposed narrow-band noise model, with arbitrarily placed side-lobe nulls, has the potential to outperform other SG filters when sinusoidal interferers of known frequency are present. The utility of these 'whitened' SG-filters is illustrated in a hypothetical pulse/peak-detection application using a test statistic that is shaped by the noise model.**

## I. Introduction

The automatic description, classification/detection, and localization of pulses in sampled time series, or peaks in discrete spectra, are fundamental operations in digital sensing systems. When the signature of the source or target is weak (e.g. dilute, distant, or covert), and the environment cluttered, additional processing/filtering is necessary to ensure satisfactory performance. Digital linear-phase finite-impulse-response (FIR) realizations of polynomial regression filters, sometimes known as Savitzky-Golay (SG) filters (named after the physical chemists who first applied them to the smoothing of spectra), are commonly used in such problems; for instance in: image-processing [1],[2], power-engineering [3],[4], and bio-medical applications [5],[6],[7],[8],[9]. SG-filters are popular and have a long history because they have: a low computational complexity, for fast realization in online systems; and a simple mathematical foundation, for the intuitive interpretation of their operation and output.

Despite these benefits, SG filters have a serious shortcoming, when viewed from a digital-signal-processing (DSP) perspective [10]; namely: the main-lobe width and side-lobe structure are not explicitly considered in the design process and as a consequence, their frequency response may be suboptimal. This then raises the question: "How does an SG-filter's frequency-response affect polynomial regression; thus if the main-lobe and side-lobes *could* be shaped arbitrarily, what *is* the optimal frequency-response of an SG filter?". These issues are considered in this paper.

SG-filters are usually used to reduce noise in low-pass smoothers or band-pass differentiators [1]-[9]. Due to their ability to perfectly match the desired frequency-response at the dc limit, they are particularly useful for analysing very low-frequency phenomena in oversampled time series. However it is also shown here, that a bank of polynomial analysers (based on SG-filters) may also be used to generate feature vectors for the detection and classification of peaks or pulses in noise. In such cases, the ability to mitigate the impact of clutter and interference (i.e. colored noise), using an *ab-initio* design procedure instead of trial and error, is highly desirable.

In Section II of this paper, an overview of some alternative (recursive and non-recursive) linear filtering techniques is provided for context, standard ways of designing (non-recursive) SG filters are summarized, then the possible utility of colored-noise models is mooted. In Section III, the structure of the FIR SG-filter is defined, the optimal filter coefficients that minimize the mean-square error (MSE) for a wide-sense stationary (WSS) noise process are solved, then some suitable colored-noise modelling procedures are presented. In Section IV, the frequency responses of various SG filters are compared and the ways in which noise parameters shape the response are discussed. It is shown that wide-band and narrow-band noise models have a different effect on the frequency response of the filter. In Section V, the behavior of the various designs are analyzed in a hypothetical pulse/peak detection application using a test-statistic, formed from a feature vector of polynomial coefficients, which is produced by a bank of SG filters. It is shown that the frequency response is a reliable predictor of performance in a variety of different (simulated) environmental-noise conditions.

## II. Context and Novelty

### A. Other Filters

When the templates in a bank of matched filters are orthogonalized, with respect to some weighting-function, the solution is an sliding instantiation of a regression problem (i.e. a so-called general-linear-model) and such an approach would also benefit from the treatment of colored-noise models considered in this paper.

LMS filters, in an adaptive line-enhancing configuration, are an ingenious solution to the signal-from-noise problem; however, they require a noise-only reference-input, which is not available in many situations; regardless, the filters described in this paper may be suitable as low-pass, high-pass, or band-pass filters for reference or input conditioning in an LMS framework.

The Kalman filter, with an Infinite Impulse Response (IIR) and augmented states to model colored-noise processes, is an optimal recursive solution to the problem of a non-stationary (e.g. polynomial) signal in sequentially correlated (i.e. colored) noise; however, solutions are no longer optimal when noise is non-Gaussian; furthermore, after filter initialization, convergence to steady-state may be rapid, making gain computations an un-necessary overhead thereafter, and solutions may not even be possible if the state covariance matrix becomes ill conditioned [33]; therefore, it is unsuitable in some applications.

The FIR Wiener filter is an optimal solution to the minimized MSE problem when the signal and the noise are both WSS processes. With the extension to incorporate non-stationary signals described by Johnson (see [11] & [12]), the solution is also optimal for the problem considered here – non-stationary (polynomial) signals in noise that is WSS, correlated, and not-necessarily Gaussian. Johnson's formulation is an alternative (more general) way of deriving the coefficients of an SG filter. The fact that colored-noise models are not utilized in standard SG-filter designs is probably because suitable forms of parameterizable noise models and their potential utility have not been described elsewhere in the literature.

### B. Standard SG Filters

SG-filters are used widely in science and engineering because the filter coefficients of recursive and non-recursive FIR realizations are readily derived from the desired polynomial degree and optionally, the derivative order [13],[14],[15],[16],[17]. As low-degree polynomials (e.g. quadratic) are usually adequate, there are many degrees of freedom in an FIR realization and not all are required for white-noise-gain (i.e. variance) minimization. For instance, some may be allocated to set the approximate filter bandwidth (i.e. the cut-off frequency) and improve frequency selectivity (e.g. to lower side-lobes or to cancel interference); however, this flexibility is not exploited in standard design procedures. Depending on the application and interpretation, increased bandwidth decreases transient-response duration (i.e. bias), decreases the scale of analysis, and increases modelling-error tolerance thus susceptibility to unknown perturbations. For a fixed polynomial degree, the side-lobes and variance of an FIR SG-filter are lowered by increasing the filter order; however, this also decreases the bandwidth [10]. This inflexibility of standard SG-filters is evident in a number of applications where polynomials of high degree are necessary to achieve the desired bandwidth in low-variance designs [4],[8],[10]. It is recommended here that the polynomial degree should be matched to the known characteristics of the signal and not used as a means of setting the filter bandwidth or side-lobe height.

Kernel functions (e.g. Gaussian [18], Epanechnikov [19], or tri-cubic [20]) that weight the regression residuals are a convenient way of tapering the impulse response and lowering the side-lobes of the frequency response. These functions are a heuristic generalization of the weights used in weighted least-squares regression [21]. Regularization methods that penalize derivatives are another way of suppressing the high-frequency gain [1],[22],[23]. These modifications do have the desired effect; however, trial-and-error tuning may be required for the desired bandwidth as they lack frequency-domain design parameters. The relationship between Tikhonov regularization and Wiener smoothing is explored in [24]. Standard DSP window-functions (i.e. Hamming, Hann, Blackman, and Tukey) are used to lower the side-lobes and high-frequency noise response of SG-filters in [2].

### C. *Whitened SG Filters*

Sequentially correlated (or so-called 'colored') noise models are used here to influence the bandwidth and side-lobe structure of SG-filters, independently of the other design parameters (i.e. polynomial degree and filter order). Such models are routinely used to steer nulls in (FIR) beamformers and in (IIR) trackers (via state augmentation or pseudo measurements) [25]-[32],[33],[34]. When possible signal perturbations are well understood, colored-signal (or process-noise) models may also be used to shape the response of FIR [35] or IIR [33],[36],[37],[38] state-estimators. It is shown in this paper that colored-noise models are a simple way of overcoming the well-known shortcomings of standard SG-filters.

In the approach presented here, second-order (stationary, autoregressive and narrow-band) Gauss-Markov processes are parameterized to cancel narrow-band interference at arbitrary frequencies. In the first-order case, a so-called 'Nyquist' process with a single pole on the $(-1,0)$ interval of the complex $z$-plane (after discretization) and a lightly-damped oscillatory autocorrelation function, is shown to be more useful than the standard non-oscillatory exponential-model that is sometimes used in tracking (e.g. Kalman) filters [25]-[32], and in FIR tracking filters [35], because it does not overlap with the polynomial signal-model near dc. Indeed, when correlated noise *is* considered in such filters described in the literature, this non-oscillatory first-order model, with a pole on the negative (real) axis of the complex $s$-plane, for a pole on the $(0,1)$ interval of the complex $z$-plane (after discretization), is generally assumed [25]-[32]. These SG-filters, with narrow-band noise models, are the primary novel contribution of this paper.

As the order of FIR filters is not determined by the number of (signal and noise) process poles, many narrow-band models may be deployed, in principle; however, in the absence of prior information regarding the structure of the noise, the limiting case of a wide-band noise model, spanning the Nyquist frequency down to the filter's cut-off frequency, is a more robust solution, with colored-noise replacing white-noise in the gain minimization. This wide-band model yields filter coefficients that are identical to those reached via the minimization of the SG-filter's stop-band energy in [39] and similar to those reached via the maximization of pass-band energy concentration in [19]. In those frequency-domain methods, the responses are subject to derivative constraints at dc and the desired endpoint is reached without explicitly leveraging noise models. The re-casting of these (less well-known) SG-filter design techniques within a wide-band noise-modelling framework, for a more intuitive physical and statistical interpretatation, is a secondary contribution of this paper.

The filters proposed here are a non-stationary (polynomial) variant of the Wiener filter and an FIR realization of a Kalman filter (augmented by noise states) at steady-state. In the Section that follows, Johnson's derivation is used to unify SG-filter and Wiener-filter design procedures, for a polynomial-regression filter that incorporates a colored-noise model. Two forms (wide-band then narrow-band) of parameterizable colored-noise models are then presented.

## III. FORMULATION

### A. *Digital filter realization*

A non-recursive FIR digital filter of order $\mathcal{O}_M$ (i.e. the number of zeros in the complex $z$-plane) is realized by convolving the input waveform $x$, with the filter coefficients $h$, to yield the filtered output $y$, using the linear difference equation

$$y[n] = \sum_{m=-K_M}^{K_M} h[m]x[n-m] \qquad (1)$$

where $n$ is the sample index and $m$ is the shift index. To simplify notation and analysis, negative shift indices (i.e. advances) are used in (1) for non-causal zero-phase filters of odd length $M = 2K_M + 1$ with $M = \mathcal{O}_M + 1$; however, the corresponding causal form is simply reached by applying a delay of $K_M$ samples. Note that square brackets are used here to enclose the indices of uniformly sampled sequences, whereas curved brackets are used to enclose the argument of the corresponding continuous function, e.g. $x[n] = x(T_s n) = x(t)$, where $T_s$ is the sampling period. The formulation here is in principle applicable to any integer-indexed sequence of equally-spaced values, for instance: samples, pixels or cells/bins; in time, space/angle or frequency; however, a sampled time-series is assumed here.

`

*B. Solving for the filter coefficients*

Johnson derives the optimal filter coefficients in (1) by minimizing the MSE, for a waveform that is a sum of signal $\mathcal{S}$, and noise $\mathcal{N}$, components, i.e. $x(t) = \mathcal{S}(t) + \mathcal{N}(t)$, where $\mathcal{N}$ is WSS, with known autocorrelation function $R$ (i.e. not necessarily Gaussian) [11]. Although, he considers the more general case with $\mathcal{S}(t) = \mathcal{P}(t) + \mathcal{M}(t)$, where $\mathcal{P}$ is a non-stationary polynomial signal of known degree ($\mathcal{O}_L$, with $L = \mathcal{O}_L + 1$ monomial terms) and $\mathcal{M}$ is another WSS process. However, the latter term is usually neglected [25],[40], as there are more than enough aspects to model without it, thus $\mathcal{M}(t) = 0$ is also assumed here. In this case, the FIR solution is readily found using

$$\begin{bmatrix} \boldsymbol{h}_{M\times 1} \\ \boldsymbol{\xi}_{L\times 1} \end{bmatrix} = \begin{bmatrix} \boldsymbol{R}_{M\times M} & \boldsymbol{\psi} \\ \boldsymbol{\psi}^T & \boldsymbol{0}_{L\times L} \end{bmatrix}^{-1} \begin{bmatrix} \boldsymbol{0}_{M\times 1} \\ \boldsymbol{\mu}_{L\times 1} \end{bmatrix} \tag{2}$$

where: $\boldsymbol{h}[K_M - m] = h[K_M + m]$, i.e. indexing direction reversed; $\boldsymbol{\xi}$ contains the Lagrange multipliers, which are ignored; $\boldsymbol{R}$ is a symmetric Toeplitz matrix, which is derived from the auto-correlation function, i.e. $\boldsymbol{R}[m_0, m_1] = R[m]$, with $m = |m_0 - m_1|$ for a WSS noise-process; $\boldsymbol{\psi}$ is a Vandermonde matrix with the discrete monomials $\psi_l[m]$ as its columns, i.e. $\boldsymbol{\psi}[m, l] = m^l$ for $m = -K_M \ldots K_M$ and $l = 0 \ldots L - 1$; and $\boldsymbol{\mu}$ contains the so-called 'moment constraints' [25]. Note that $[\cdot]^T$ is a transpose operation. Alternative IIR problem formulations are discussed in [40],[41]; and an analogous frequency-domain form is presented in [39] – referred to there as a 'colored' SG-filter. Equation (2) is a limiting case of Johnson's more general solution in [11], with a polynomial-only signal-model and symmetric indexing for a non-causal zero-phase realization [25]. In this case, using the matrix identities provided in [11] & [12] – see equations (22)-(25) in [11] and errata in [12] – equation (2) above simplifies to the more familiar form for weighted least-squares regression

$$\boldsymbol{H}_{L\times M} = \{\boldsymbol{\psi}^T \boldsymbol{W} \boldsymbol{\psi}\}^{-1} \boldsymbol{\psi}^T \boldsymbol{W} \text{ then} \tag{3a}$$
$$\boldsymbol{h}^T = \boldsymbol{\mu}^T \boldsymbol{D}^d \boldsymbol{H} \tag{3b}$$

where $\boldsymbol{\mu}^T = [1 \quad \boldsymbol{0}_{1\times(L-1)}]$ for phase linearity, $\boldsymbol{D}^d$ (an $L \times L$ matrix) is the $d$th-order derivative operator i.e.

$$\boldsymbol{D} = \begin{bmatrix} 0 & 1 & 0 & 0 & 0 & \ddots \\ 0 & 0 & 2 & 0 & \ddots & 0 \\ 0 & 0 & 0 & \ddots & 0 & 0 \\ 0 & 0 & \ddots & 0 & L-2 & 0 \\ 0 & \ddots & 0 & 0 & 0 & L-1 \\ \ddots & 0 & 0 & 0 & 0 & 0 \end{bmatrix}_{L\times L} \text{ and } \boldsymbol{D}^d = \boldsymbol{I} \underbrace{\times \boldsymbol{D}}_{d} \tag{3c}$$

(e.g. for $L = 3$, $d = 2$ and $T_s = 1$, $\boldsymbol{D}^2$ is a matrix of zeros, with 2 in the top-right corner) and $\boldsymbol{W} = \boldsymbol{R}^{-1}$ (an $M \times M$ matrix) transforms the input in a manner that whitens the input noise. Note that when $\boldsymbol{W}$ is diagonal (i.e. not derived from the autocorrelation function of a WSS process) the transform is a simple element-wise scaling operation, with diagonal coefficients equal to the kernel $w[m]$. Kernel selection is somewhat arbitrary and the fields of DSP and statistics offer up many possible alternatives [19]. If the kernel is everywhere positive then it may be interpreted as a weight and normalized accordingly, then used to evaluate expected errors and moment properties, from the fitted residual and the polynomial coefficients, respectively. Alternatively, a parameterizable noise model provides a means of defining all elements of $\boldsymbol{W}$ so that more degrees of freedom are utilized for optimal smoothing in any given application or environment.

*C. Colored-noise models*

An odd-symmetric FIR-filter has $N_M$ unknown coefficients to determine, thus $N_M$ conjugate zeros to place in the complex $z$-plane, where $N_M = K_M + 1$, with $K_M$ being half the number of coefficients at non-zero shifts; however, $N_L = K_L + 1$ degrees of freedom are required to preserve moments up to $\mathcal{O}_L$th degree in the $n$-domain thus dc-flatness up to $\mathcal{O}_L$th degree in the $\omega$-domain, where $K_L$ is half the number of non-zero monomials, such that $L = 2K_L + 1$, and where $\omega$ is the angular frequency (radians/sample). Therefore, there are $N_K - N_L$ degrees of freedom remaining that may be used for zero (or null) placement in the $z$-plane (or $\omega$-domain). Using $\boldsymbol{R}$ in (2) and (3) ensures that the poles of the noise process are (approximately) canceled by zeros of the filter.

When noise processes are largely unknown, a wide-band (WB) model, with unity spectral density $S_{\mathrm{WB}}(\omega)$ over $\omega_c \leq$

`

$|\omega| \leq \omega_d$ and zero elsewhere, is appropriate. In this case, the noise is simply defined as being 'everything that is definitely not signal' thus $0 < \omega_c < \pi$ and $\omega_d = \pi$. Note that $\omega_c = \Omega_c T_s$, where $\Omega_c$ is the cut-off frequency of the noise process (radians/second). The Wiener–Khinchin theorem then yields

$$R_{\mathrm{WB}}[m] = \mathcal{R}_{\omega_d}[m] - \mathcal{R}_{\omega_c}[m] \text{ where} \tag{4a}$$

$$\mathcal{R}_{\omega}[m] = \begin{cases} 2\omega, & m = 0 \\ 2\sin(\omega m)/m, & m \neq 0 \end{cases}. \tag{4b}$$

When $\boldsymbol{R}$ is formed using the elements defined in (4), then inverted to form $\boldsymbol{W}$ in (3), it yields the band-limited SG-filters described in [19] and [39].

However, when the frequencies of interferers are known *a-priori*, a narrow-band (NB) model is more appropriate. The causal part of the proposed NB model of the noise process is defined in the complex $s$-plane as a sum of $N_{\mathrm{NB}}$ second-order terms

$$S_{\mathrm{NB}}(s) = \sum_{k=0}^{N_{\mathrm{NB}}-1}\{1/(s - s_k) + 1/(s - s_k^*)\} \tag{5}$$

where $N_{\mathrm{NB}} \leq N_K - N_L$ and $[\cdot]^*$ denotes complex conjugation. The process poles are defined using $s_k = \sigma_{\mathrm{NB}} + i\Omega_k$, where $i = \sqrt{-1}$, $\Omega$ is the natural frequency (radians/second) and $\sigma_{\mathrm{NB}} < 0$ for causal stability. Note that only the shape of the noise spectral-density is important because the inverse operation in (3a) removes any scaling factors and takes care of normalization. Applying the inverse Laplace transform and combing with the non-causal part yields

$$R_{\mathrm{NB}}(t) = \sum_{k=0}^{N_{\mathrm{NB}}-1} e^{\sigma_{\mathrm{NB}}|t|}\cos(\Omega_k t) \tag{6a}$$

which is then sampled (i.e. discretized) to yield

$$R_{\mathrm{NB}}[m] = \sum_{k=0}^{N_{\mathrm{NB}}-1} \rho^{-|m|}\cos(\omega_k m) \tag{6b}$$

where: $\omega_k = \Omega_k T_s$, i.e. the argument or natural frequency of the $k$th conjugate pole pair of the noise-process in the $z$-plane (in radians/sample); $\rho = e^{1/\lambda_{\mathrm{NB}}}$, i.e. the magnitude of the noise-process poles in the $z$-plane; $\lambda_{\mathrm{NB}} = 1/\sigma_{\mathrm{NB}} T_s$, i.e. a scale parameter (in samples). For FIR filter design, (6b) is truncated at $\pm K_M$. When $N_{\mathrm{NB}} < N_K - N_L$ , surplus degrees of freedom are deployed to minimize the white-noise gain (WNG) when (2) or (3) is solved. The achieved WNG is evaluated using either $h[m]$, or the frequency response $H(\omega)$, of the filter – as a consequence of Parseval's theorem – using

$$\mathrm{WNG} = \sum_{m=-K_M}^{K_M}|h[m]|^2 = \frac{1}{2\pi}\int_{-\pi}^{\pi}|H(\omega)|^2 d\omega. \tag{7}$$

Less-than-unity and greater-than-unity values of WNG are indicative of white-noise attenuation and amplification, respectively.

## IV. Response Analysis and Parameterization

For an FIR SG-filter of $O_M$-th order and an input comprised of an $O_L$-th degree polynomial plus zero-mean white-noise with a variance of $\sigma_x^2$, the expected squared-error at steady-state (i.e. after processing $M$ samples) is $\sigma_e^2 = \mathrm{WNG} \times \sigma_x^2$, [33],[41]. Similarly, when an additive interferer $\mathcal{N}[n] = \sin(\omega_x n + \phi_x)$ replaces the white noise, $\sigma_e^2 = |H(\omega_x)|^2$ [33]; thus the filter error may be read directly from the magnitude response of the filter.

The magnitude responses $|H(\omega)|^2$, of various FIR SG-filters, with $M = 17$ and $L = 5$, are compared in Fig. 1 and discussed in the remainder of this section. The filters were designed using different $\boldsymbol{W}$ matrices; all noise processes were discretized using $T_s = 1$; other details are provided below. This particular sequence of filters (Filters A through G) was chosen to illustrate a conceptual progression of model sophistication and design complexity.

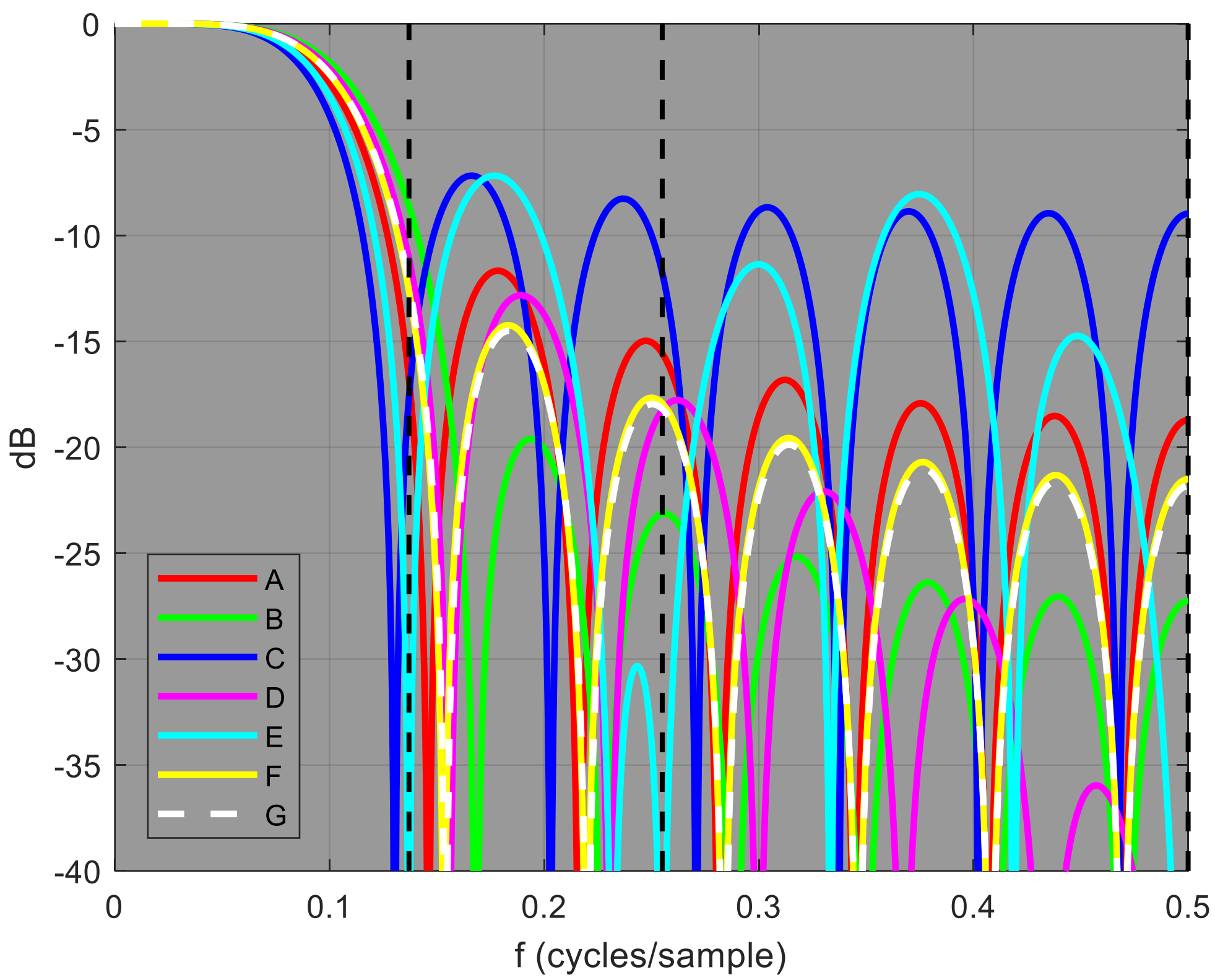

**Fig. 1**. Magnitude response of various FIR SG-Filters ($f = \omega/2\pi$).

*A. Filter A*

Filter A (WNG = 0.2103) is a standard SG-filter with $\boldsymbol{W} = \boldsymbol{I}$. The cut-off frequency $\omega_c$, used to design Filters B, E, F & G was set using $\omega_c = 0.75\omega_\Delta$, where $\omega_\Delta$ is the frequency of Filter A's first null. For Filter A, $\omega_\Delta$ is determined post-design; it is equal to the argument of the zero of the filter's transfer function $H(z)$, that is closest to the unit circle and closest to dc. For the example considered in this section $\omega_\Delta = 0.9179$. This heuristic determination of $\omega_c$ is an approximation of the -3 dB gain frequency and it ensures that all filters have similar bandwidth for ease of comparison. Filter A has the lowest WNG of all filters, because all degrees of freedom are used to minimize noise variance. In the absence of a colored-noise model, as $M$ increases and $L$ decreases: the width of the main-lobe and the height of the side-lobes decrease, for a lower WNG, thus increased smoothness of $y$ [10].

*B. Filter B*

Filter B (WNG = 0.2204), has a symmetric Gaussian kernel along the diagonal of $\boldsymbol{W}$, with all other elements equal to zero [18]. The scale $\lambda_n$, or standard deviation of the Gaussian was determined by setting the three-sigma point of the kernel's frequency-response (also approximately Gaussian) equal to $\omega_c$, i.e. $\lambda_\omega = \omega_c/3 = 1/\lambda_n$. Relative to Filter A, this filter has lower side-lobes and a wider main-lobe.

*C. Filter C*

Filter C (WNG = 0.2374) uses a first-order low-pass Gauss-Markov noise model to form $\boldsymbol{W}$, with $\rho = \exp(-1/\lambda_n)$. Although this model is popular (e.g. [25]-[32]), and reasonable when $T_s$ is small enough to reveal temporally correlated measurement errors, it is not particularly useful because it is unreasonable to expect the filter to separate a low-frequency signal from low-frequency noise. As a result, the filter's side-lobes and WNG are the highest of all filters.

*D. Filter D*

Filter D (WNG = 0.2166) also uses a first-order Gauss-Markov noise model; however in this case, $\rho = -\exp(\sigma_{\text{NB}}T_s)$, where $\sigma_{\text{NB}}$ is equal to a negative value that is: small enough to place a filter zero just inside the unit circle for a deep notch at $\omega = \pi$; yet large enough to ensure that $\boldsymbol{R}$ is not ill conditioned. A value of $\sigma_{\text{NB}} = -1.0 \times 10^{-6}$ was used for the NB models in Filters D & E. This Nyquist noise model was also used in the IIR tracking filters of [33]; note that $\sigma_{\text{NB}} = 0$ was used there, because explicit matrix inversions are not required in those recursive state-observers. The dependence of the filter response on the location of the real part of the noise-process pole in the complex $s$-plane (i.e. $\sigma_{\mathrm{NB}}$) is illustrated in Fig. 2.

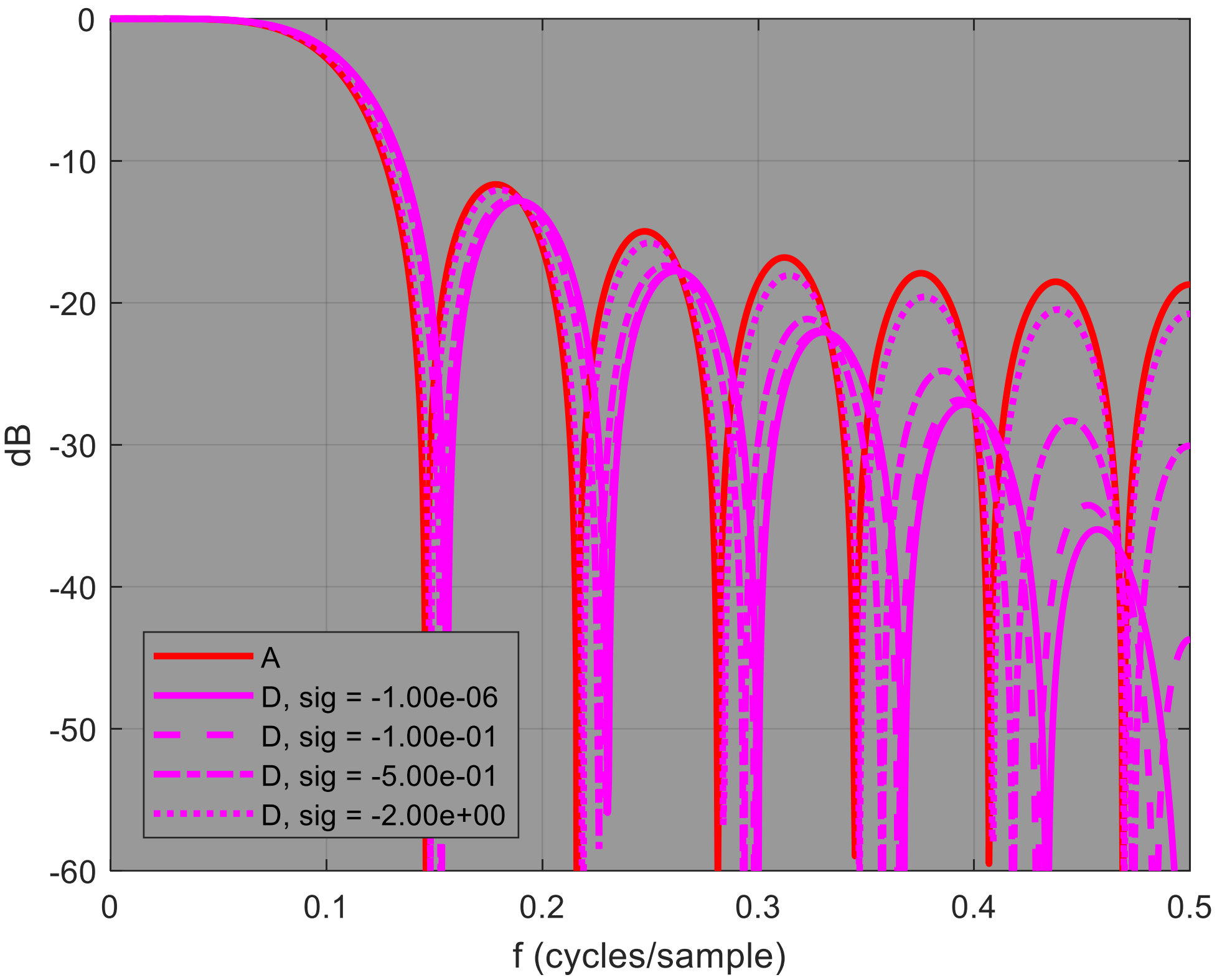

**Fig. 2**. Magnitude response of Filter D for various values of $\sigma_{\mathrm{NB}}$ in a (first-order) narrow-band model. Response of Filter A is provided as a reference. The height of the Nyquist side-lobe (at $f = 0.5$) is lowered until it effectively becomes a notch, as $\sigma_{\mathrm{NB}} \to 0$ (from the left, in the $s$-plane), such that $\rho \to -1$ (from the right), which shifts one or more filter zeros towards the unit circle (in the $z$-plane).

### *E. Filter E*

Filter E (WNG = 0.2290) uses the proposed NB noise model with $N_{\mathrm{NB}} = 3$. The natural frequencies of the model may be chosen arbitrarily to best suit a given noise environment; however, a simple null pattern was generated here, using $\Omega_k \in \{0.8608, 1.6022, 3.1416\}$, i.e. the dashed black lines in Fig. 1, $f_k \in \{0.137, 0.255, 0.5\}$. Note that for all other filters, the lowest NB frequency is at the edge of the main-lobe (Filter C is an exception, by chance) and the other NB frequencies are near the middle of a side-lobe (Filter D is an exception, by design).

### *F. Filters F & G*

Filters F & G (WNG = 0.2119 & 0.2122) both use a WB noise model; however for Filter G, the solution is iteratively refined using the procedure described in [19], to maximize energy concentration in the passband (i.e. $|\omega| \le \omega_c$). The effect is barely noticeable in Fig. 1, with a slight decrease in the side-lobe height, at the expense of a slight increase in the main-lobe width and WNG. The dependence of the filter response on the noise cut-off frequency ($\omega_c$) is illustrated in Fig. 3.

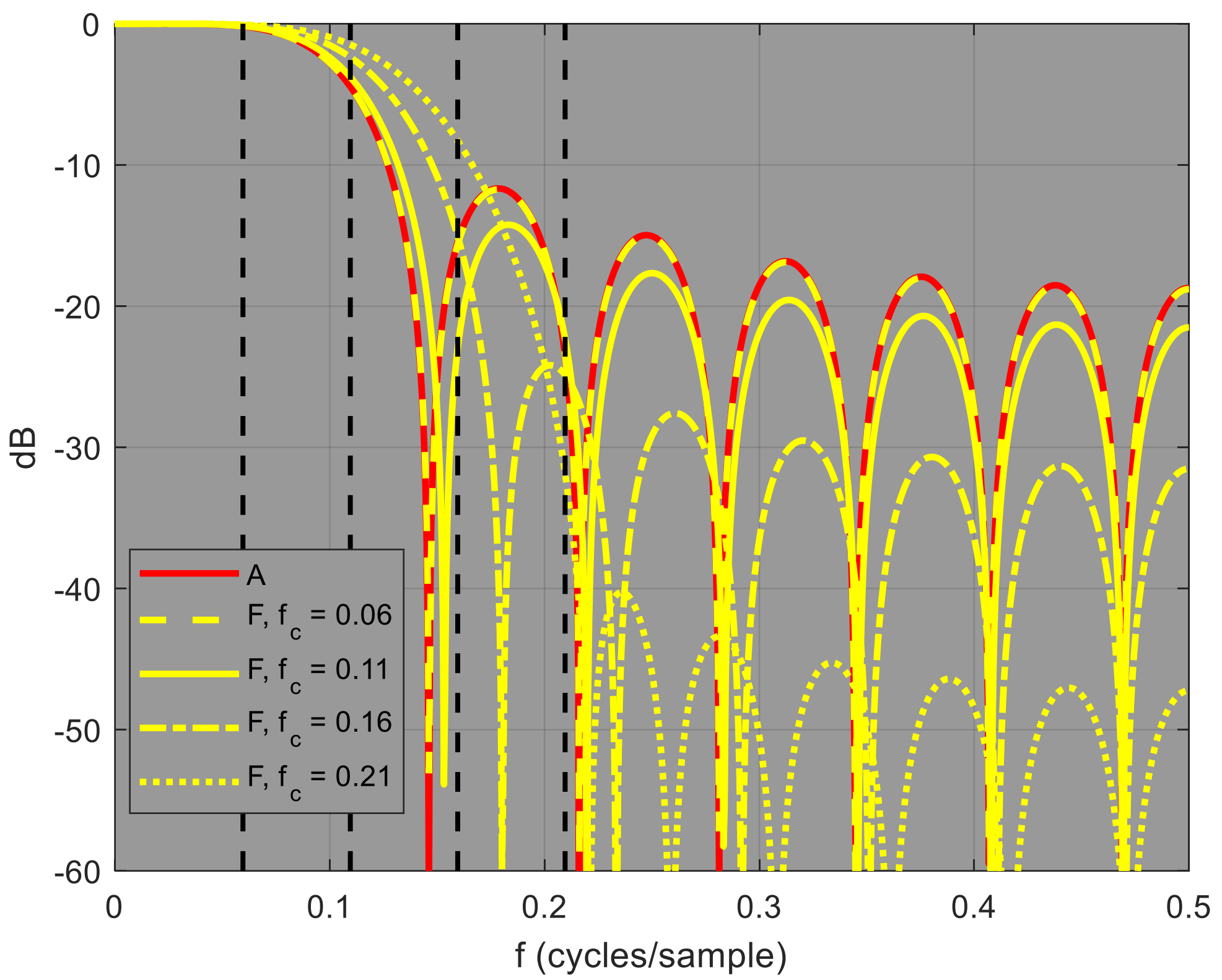

**Fig. 3**. Magnitude response of Filter F for various wide-band cut-off frequencies ($f_c = \omega_c/2\pi$, dashed black lines). Response of Filter A is provided as a reference. The main-lobe width increases, and side-lobe height decreases, as $f_c$ increases. The response of Filter F approaches that of Filter A as $f_c \to 0$, i.e. as the color of the noise fades to white.

## V. Application and Illustration

A noise-model formalism maintains the connection between the desired response in the $\omega$-domain and the regression framework in the $n$-domain via a simple sequence of transformations ($S \to R \to \boldsymbol{W}$), which supports the statistical analysis of non-local properties (i.e. over the $\pm K_M$ interval) using the filter outputs in a classifier or detector (i.e. a binary classifier). A simple pulse detector, framed as a sliding hypothesis test, is used as an example in this section to illustrate the possibilities and to demonstrate the benefit of parameterized noise models. A bank of SG filters generates estimates of polynomial coefficients (a local feature vector) that are then combined to yield the test statistic at each sample. A detection event is declared when the test statistic exceeds a specified threshold.

### A. *An SG filter-bank*

For a block of waveform samples, centered on $n$ and contained in the $M \times 1$ data-vector $\boldsymbol{x}$, estimates of the monomial coefficients in the $L \times 1$ feature-vector $\boldsymbol{\alpha}$, are simply determined via the *analysis* operation in (3a) using $\hat{\boldsymbol{\alpha}} = \boldsymbol{H}\boldsymbol{x}$, realized as an FIR filter-bank, with the coefficients of the $l$th sub-filter taken from the $l$th row of $\boldsymbol{H}$. An estimate of the (polynomial) signal over this $[n-K, n+K]$ interval may then be reconstructed using the *synthesis* operation $\hat{\boldsymbol{\mathcal{S}}} = \boldsymbol{\psi}\hat{\boldsymbol{\alpha}}$. An analysis operation is applied to generate local estimates of polynomial coefficients $\hat{\boldsymbol{\alpha}}$, for the test statistic $Z$, that is used by the detector; the synthesis operation (with $d = 0$ thus $\boldsymbol{D}^d = \boldsymbol{I}$) is only required to reconstruct the smoothed signal for visualization purposes. This is done by evaluating the fitted polynomial at the center of the analysis window, in the linear-phase case (for a FIR filter with a group delay of $K$ samples), using (3b). Details of this analysis/synthesis structure for both FIR and IIR filters with an arbitrary group delay (i.e. not necessarily linear-phase) are described elsewhere [41].

### B. *A pulse-detection test-statistic*

For the detection of simple quadratic pulses, the $l = 2$ monomial term is the most important signal component. Furthermore, it is assumed that large dominant pulses and smaller less-obvious pulses are of equal importance; therefore, the $l = 0$ (i.e. constant) term is ignored. It is also assumed that odd terms (e.g. $l = 1$ or $l = 3$, i.e. linear or cubic) are an indication of non-maximal features (e.g. steps, ramps, or non-centered pulses) and that high-order even terms (e.g. $l = 4$ and beyond) are indicative of other non-pulse features (e.g. impulses). For this purpose, the following heuristic test-statistic is proposed:

$$Z[n] = \frac{P_1}{P_0} = \frac{\hat{\boldsymbol{s}}_1^T \boldsymbol{W} \hat{\boldsymbol{s}}_1}{\hat{\boldsymbol{s}}_0^T \boldsymbol{W} \hat{\boldsymbol{s}}_0} = \frac{\hat{\boldsymbol{\alpha}}_1^T \boldsymbol{\psi}^T \boldsymbol{W} \boldsymbol{\psi} \hat{\boldsymbol{\alpha}}_1}{\hat{\boldsymbol{\alpha}}_0^T \boldsymbol{\psi}^T \boldsymbol{W} \boldsymbol{\psi} \hat{\boldsymbol{\alpha}}_0} = \frac{\hat{\boldsymbol{\alpha}}_1^T \boldsymbol{U}^T \boldsymbol{\varphi}^T \boldsymbol{W} \boldsymbol{\varphi} \boldsymbol{U} \hat{\boldsymbol{\alpha}}_1}{\hat{\boldsymbol{\alpha}}_0^T \boldsymbol{U}^T \boldsymbol{\varphi}^T \boldsymbol{W} \boldsymbol{\varphi} \boldsymbol{U} \hat{\boldsymbol{\alpha}}_0} = \frac{\hat{\boldsymbol{\beta}}_1^T \hat{\boldsymbol{\beta}}_1}{\hat{\boldsymbol{\beta}}_0^T \hat{\boldsymbol{\beta}}_0}. \tag{8}$$

It is conceptually similar to a test statistic that has been used elsewhere to analyze oriented structure in 2-D signals (e.g. images of fingerprints [42]); with additional 'clutter' terms included here, in the denominator, for improved discrimination of 1-D quadratic shapes.

The $Z$ statistic is a dimensionless ratio of filtered autocorrelations ($P_1$ and $P_0$), accumulated over the filter's finite analysis window where: the $[\cdot]_1$ subscript corresponds to the foreground-signal (i.e. pulse) hypothesis with all coefficients in $\hat{\boldsymbol{\alpha}}$ zeroed, except for the $l = 2$ term; and the $[\cdot]_0$ subscript corresponds to the background-signal (i.e. clutter) hypothesis, with the $l = 0$ and $l = 2$ coefficients in $\hat{\boldsymbol{\alpha}}$ set to zero, to remove any constant (i.e. dc offset) and quadratic contributions, respectively. Evaluation of $Z$ is greatly simplified by transforming the monomial basis-set into an ortho-normal basis-set, such that $\boldsymbol{\varphi}^T \boldsymbol{W} \boldsymbol{\varphi} = \boldsymbol{I}_{L\times L}$ where $\boldsymbol{I}$ is the identity matrix and $\boldsymbol{\varphi}$ has the discrete ortho-normal basis-functions as its columns. The required transformation is found via a Cholesky decomposition such that $\boldsymbol{\psi}^T \boldsymbol{W} \boldsymbol{\psi} = \boldsymbol{U}^T \boldsymbol{U}$, thus $\boldsymbol{\psi} = \boldsymbol{\varphi} \boldsymbol{U}$ and $\hat{\boldsymbol{\beta}} = \boldsymbol{U} \hat{\boldsymbol{\alpha}}$, where $\boldsymbol{U}$ is an upper-triangular ($L \times L$) matrix. A detection is declared at $n$ when $Z[n]$ exceeds a specified detection threshold $\gamma_Z$. The scale and form of the desired transients is assumed to be known *a priori*; therefore, adaptive-order algorithms (e.g. [23]) are not considered here.

As the test statistic is the local signal power over the local clutter power, it may be interpreted as signal-to-noise ratio (SNR). In practice, a small constant value should be added to the denominator of (8), to protect against division-by-zero errors and very large values of $Z$ in low-noise environments. A value of 0.001 was used here; and to ensure that it had a similar affect for all filters (i.e. reduced sensitivity), the $\boldsymbol{W}$ matrices were scaled so that the maximum element was equal to unity.

### C. *Analysis of detection performance*

The behavior of the filters described in the previous section and the resulting detectors are illustrated in Figs. 4-7 for simulated (pseudo-randomly generated) inputs with $T_s = 1$ (s). The upper subplot shows $y[n]$ for all filters (for $d = 0$, i.e. the smoothed or low-pass filtered output) on the synthetic input sequence $x[n]$ (in black), which contains from left to right: quadratic pulses of small, medium and large duration (9, 17 & 33 samples, respectively), followed by medium-sized square and saw-tooth pulses. The sign of each pulse in the simulated sequence alternates to verify that the test statistic is sign independent.

The pulse train was added to a 2 mV dc offset and a low-frequency ($\Omega_{\mathrm{NB}} = 0.0084$) 'background' sinusoid with a magnitude of 5 mV and a random phase offset. Two white-noise and two colored-noise scenarios were considered. Randomly-generated zero-mean Gaussian-noise, with $\sigma_x = 0.1$ mV and 0.2 mV, was added in the 'low' and 'high' white-noise scenarios, respectively (see Fig. 4 & 5); $\sigma_x = 0.2$ mV was also used in both colored-noise scenarios. Two sinusoids with $\Omega_{\mathrm{NB}} = 0.8608$ & $\Omega_{\mathrm{NB}} = 1.6022$, i.e. matched to two of the three frequencies used to design Filter E, were added in the first 'matched' colored-noise scenario (see Fig. 6). Two sinusoids with $\Omega_{\mathrm{NB}} = 0.9469$ & $\Omega_{\mathrm{NB}} = 1.7624$, i.e. $f_{\mathrm{NB}} = 0.1507$ & $f_{\mathrm{NB}} = 0.2805$, were added in the second 'mismatched' colored-noise scenario (see Fig. 7). In both colored-noise scenarios, these sinusoidal 'interferers' had a magnitude of 5 mV and a random phase offset. The input waveform is delayed by $K_M$ samples in the upper subplots to compensate for the group delay of the filters – so that $x$ & $y$ are approximately aligned to aid visualization. The detection test-statistic $Z$, for Filters A-F is shown in the lower subplot (see Fig. 1 for legend). This metric was not computed for Filter G, because $\boldsymbol{\psi}^T \boldsymbol{W} \boldsymbol{\psi}$ is not necessarily positive definite, which is required for a Cholesky decomposition.

One thousand instantiations of each scenario were randomly generated and receiver operating characteristic (ROC) curves were generated by adjusting the detection threshold in 0.1 dB steps from 0 dB to 100 dB. The probability of detecting the pulse to which the filter is matched ($P_D$), i.e. the medium quadratic pulse; and the probability of false detection ($P_F$), i.e. on another pulse or some other artifact (e.g. noise or interference), were computed for each threshold applied. Differences between the ROC curves for the various filters, in the low white-noise scenario, are barely discernable (not shown); a detail of the ROC curve for the high white-noise scenario is shown in Fig. 8. Not surprisingly, because it has the highest WNG of all filters, Filter C is the worst detector in the high white-noise scenario, as it mostly yields the lowest $P_D$ for a given $P_F$. The ROC curves for the matched and mismatched colored-noise

`
scenarios are shown in Figs. 9 & 10, respectively. Filter E is the best detector in the matched scenario (see Fig. 9), and one of the worst detectors in the mismatched scenario (see Fig. 10), because it has nulls at the interfering frequencies in the former case (by design), and one of the highest gains, due to the position of its side-lobes, in the latter case (see Fig. 1). Similarly, Filters A & F have the best detection performance in the mismatched scenario because they have nulls near the interfering frequencies (by chance).

The first random instantiation of each scenario is plotted in Figs. 4-7 and a detection threshold of $\gamma_Z = 20$ dB was applied in all cases (i.e. the dotted horizontal line in the lower subplots); detections produced by the filters are marked with '×' tokens. These plots are provided and discussed below, to visualize the scenarios, to illustrate the procedure used to generate the ROC curves, and to account for the detection behavior of the various filters.

In the absence of noise, the detector should detect a quadratic pulse with a duration that is at least as large as $M$; however, when noise and interference are present, for a smaller window, the lower curvature of larger pulses may mean that there is insufficient foreground power $P_1$, relative to the background power $P_0$, which suppresses the SNR. Furthermore, the low-frequency background sinusoid contributes to an inflated denominator via the $l = 1$ signal term which may also reduce the SNR slightly for some quadratic pulses, where the background gradient is large.

In the low white-noise scenario (see Fig. 4), all filters correctly detect the medium quadratic pulse to which they are tuned (Filters E and F also detect the larger quadratic pulse). In the high white-noise scenario (see Fig. 5), all filters detect the medium quadratic pulse only (with the exception of Filter C, which produces no detections for the chosen threshold). In both white-noise scenarios, detections on the non-quadratic pulses are correctly suppressed, due to the contribution of non-quadratic signal terms which make $P_0$ large. Note that when longer SG filters with $M = 33$ are used, all filters detect the large quadratic pulse only, in both white-noise scenarios.

In the matched colored-noise scenario (see Fig. 6), only Filter E functions correctly (for the chosen threshold) because it whitens the colored-noise. For all other filters, the interference yields spurious polynomial coefficients, which may produce false alarms (e.g. Filter F in this case) or cause missed detections when they are combined to form $Z$ (i.e. the SNR) in the lower subplot. On signal 'synthesis', these waveform 'analysis' errors then induce oscillations in the smoothed output, which are clearly evident in the upper subplot. As mentioned at the beginning of Section IV, the magnitude of these oscillatory errors is determined by the magnitude of the frequency response at the interference frequency (see Fig. 1). In the mismatched colored-noise scenario (see Fig. 7), all filters produce no detections.

The analysis in this section confirms that, in addition to smoothing and (derivative) state-estimation, SG-filters may also be useful in binary classification (i.e. detection) problems, where low-order polynomials are a reasonable description of signal features, e.g. for simple transients, pulses or peaks. In this role, the incorporation of a narrow-band colored-noise model has the potential to improve performance, when the frequencies of interferers are known. Note that it *is* possible to adequately attenuate wide-band colored noise (at the expense of white noise) using an arbitrary window function with low side-lobes (e.g. via the Gaussian taper used in Filter B); however, a wide-band noise-model (i.e. Filter F) provides a simple framework to optimally balance the main-lobe width and side-lobe height of the SG filter in a least-squared-error sense (see Fig. 3).

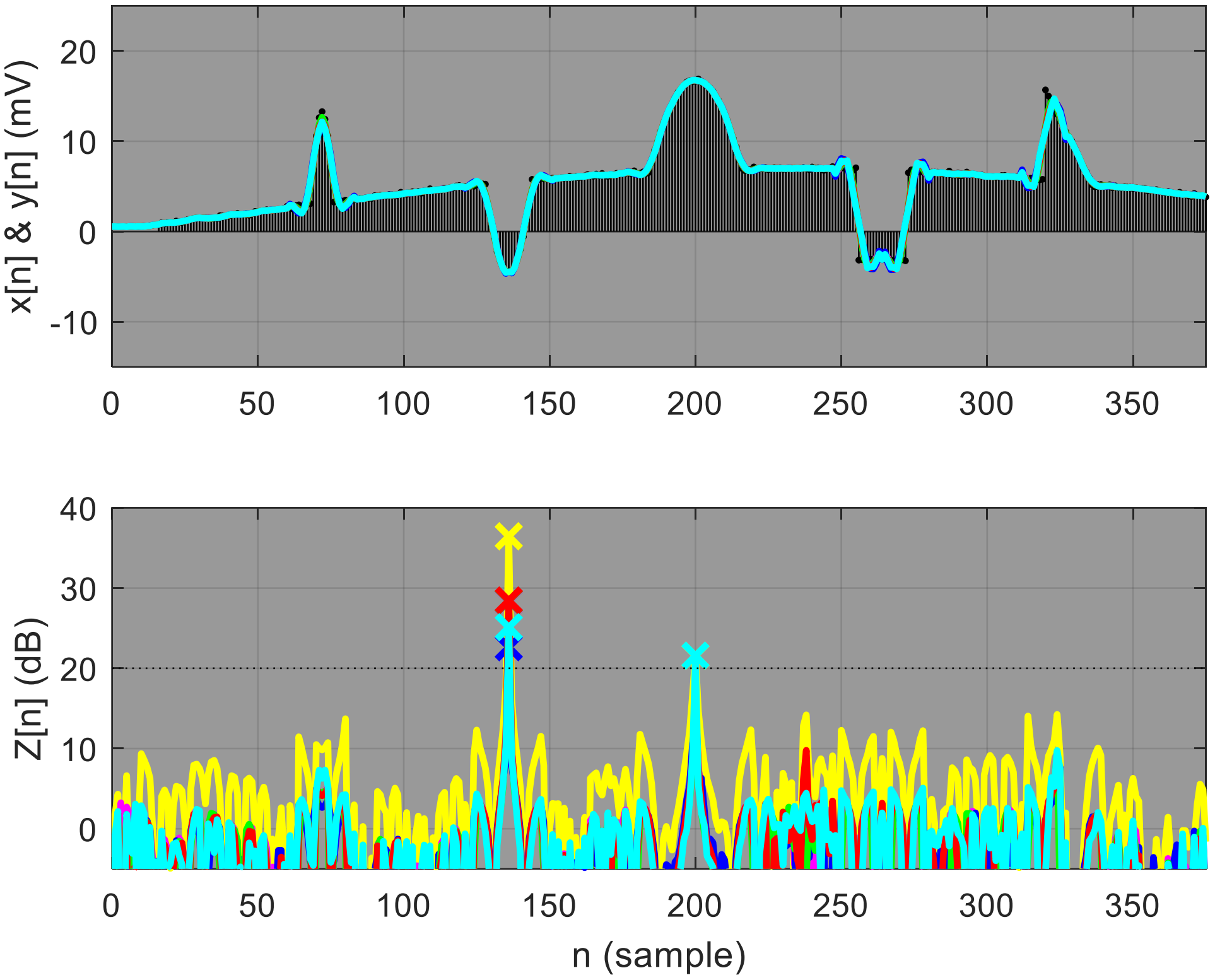

**Fig. 4**. Low white-noise scenario.

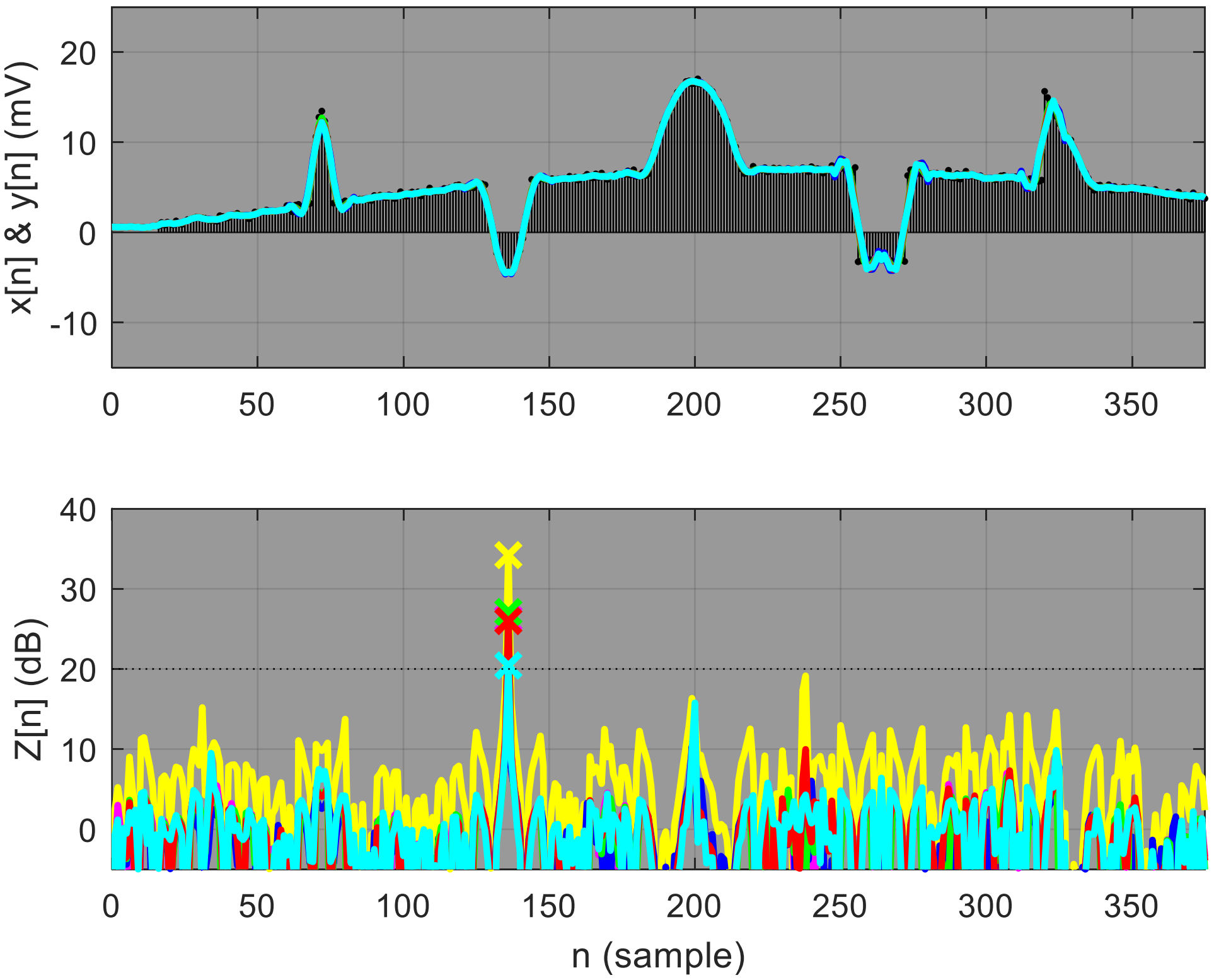

**Fig. 5**. High white-noise scenario.

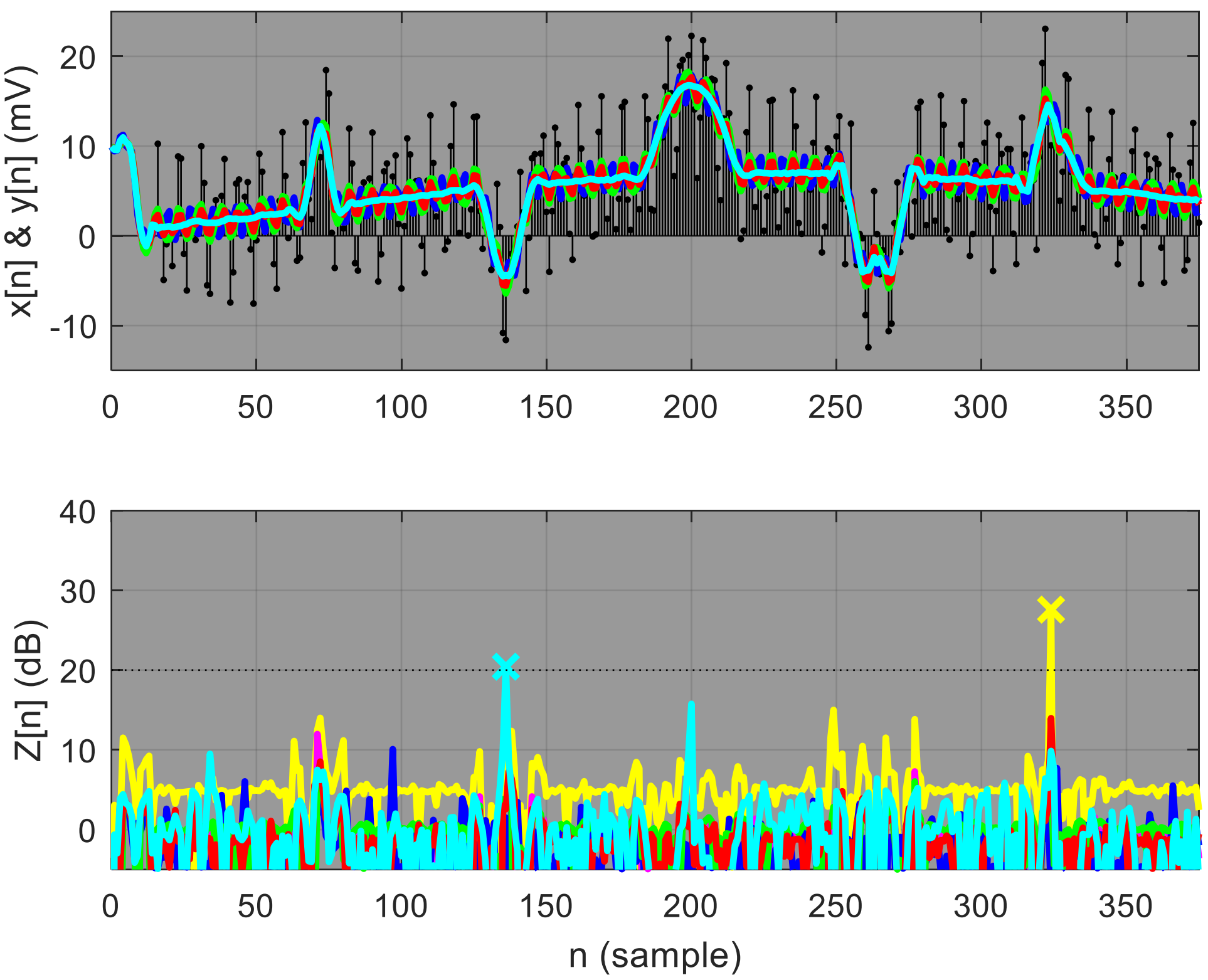

**Fig. 6**. Matched colored-noise scenario.

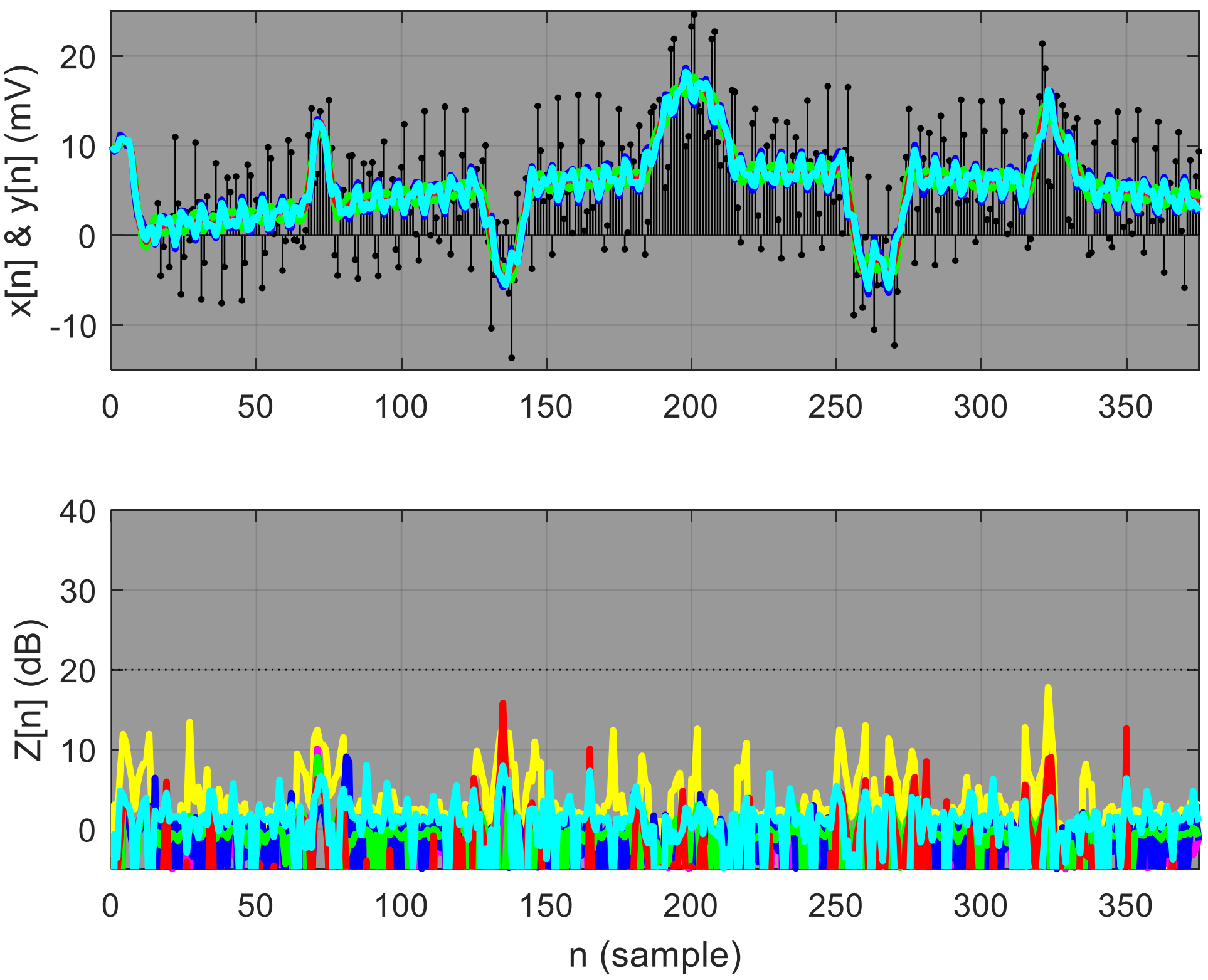

**Fig. 7**. Mismatched colored-noise scenario.

`

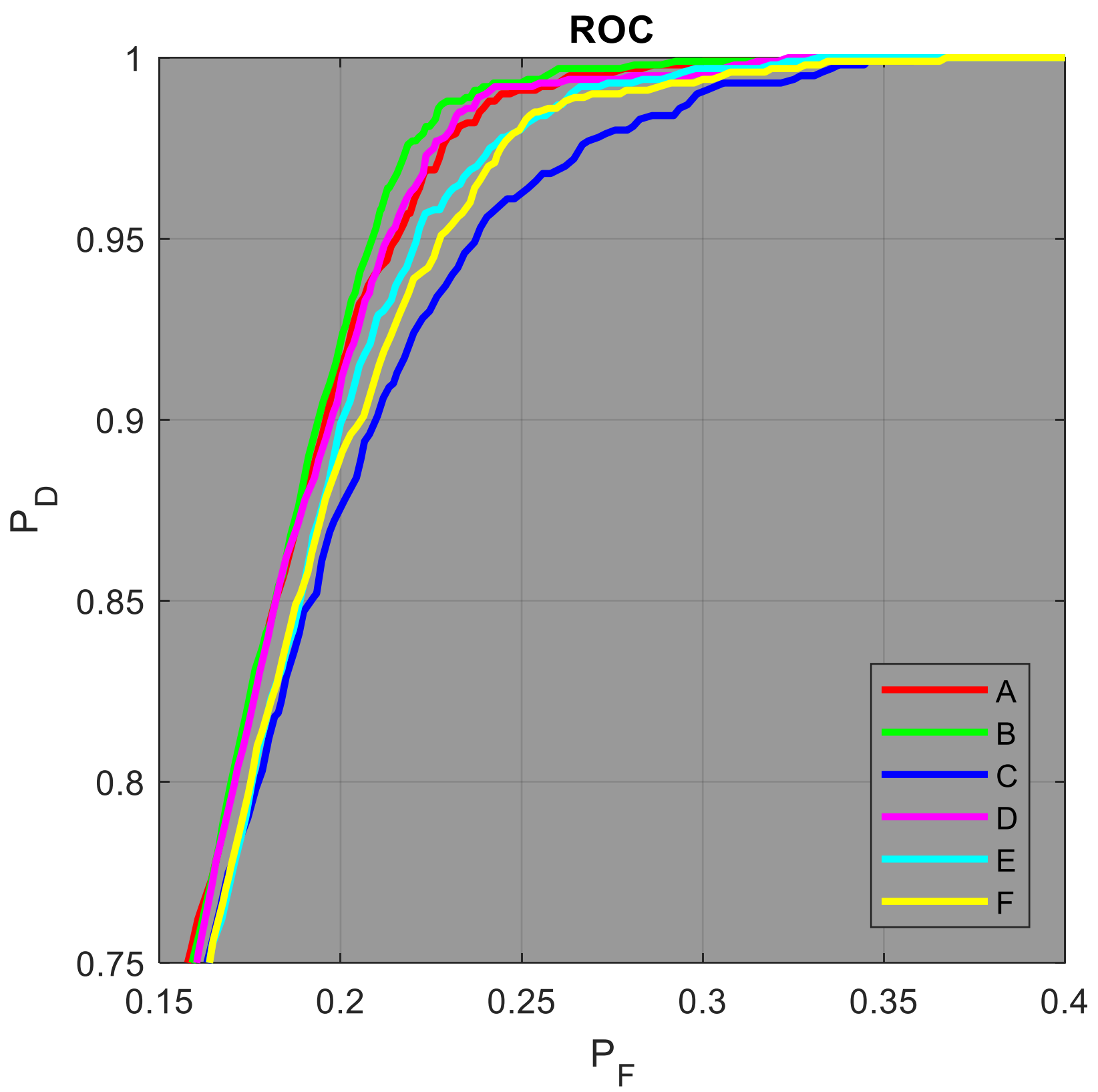

**Fig. 8**. ROC curve for the high white-noise scenario. Detail of 'knee' shown.

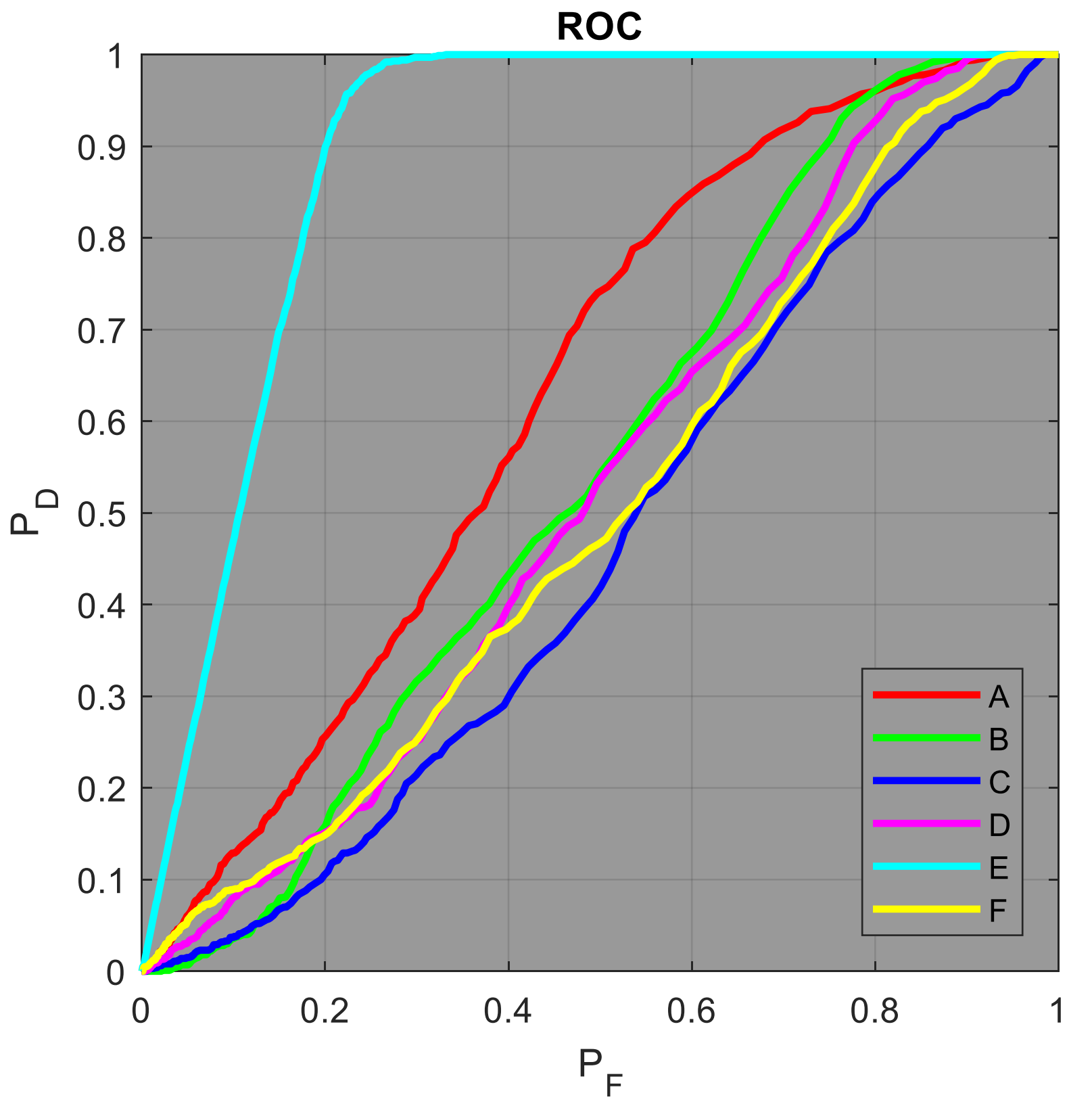

**Fig. 9**. ROC curve for the matched colored-noise scenario.

`

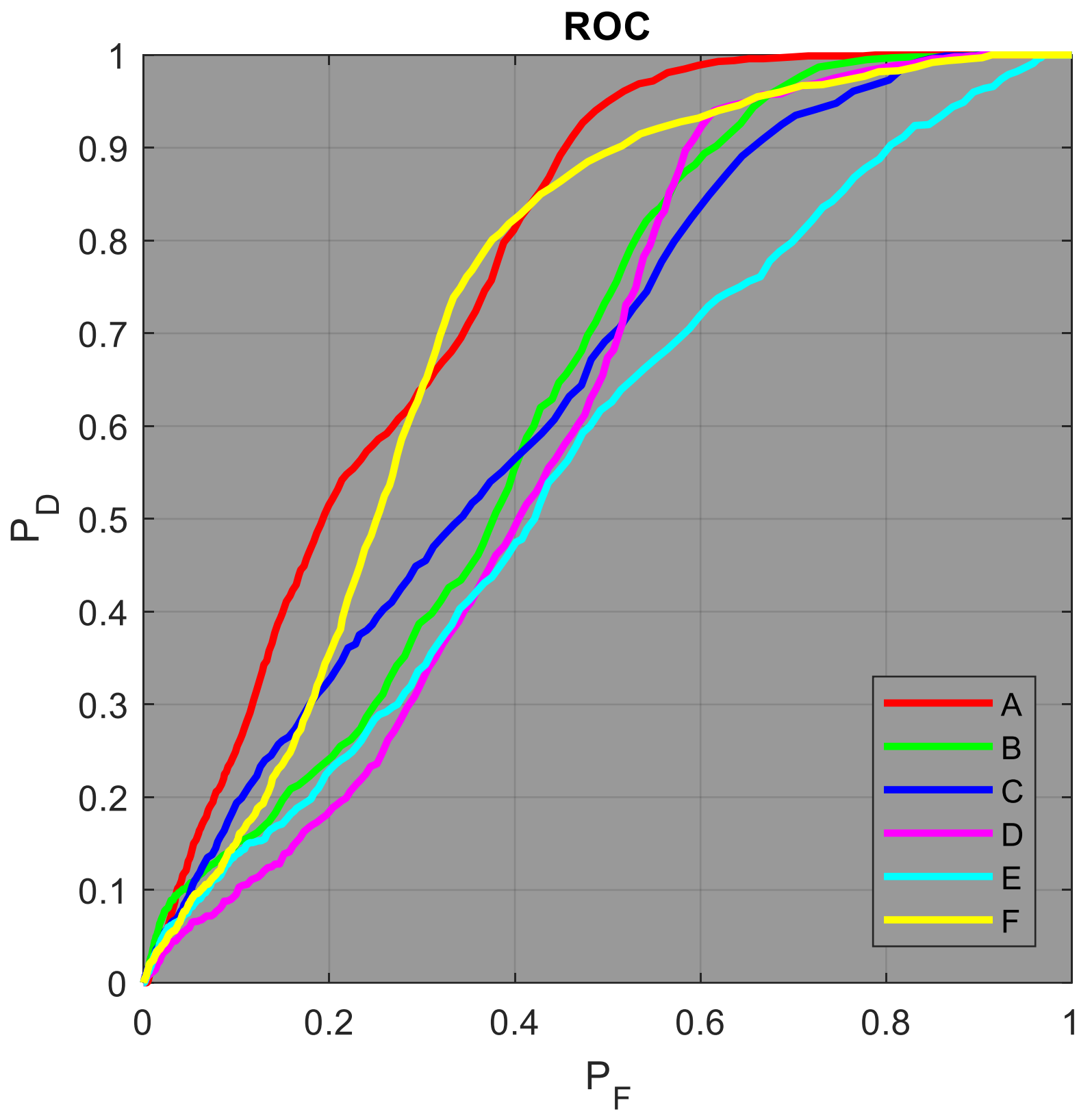

**Fig. 10**. ROC curve for the mismatched colored-noise scenario.

## VI. Conclusion

The Wiener filter is biased for non-stationary (e.g. polynomial) signals and the Kalman filter is suboptimal for non-Gaussian noise. When the prior specification and online manipulation of covariance matrices is infeasible, SS-KF realizations with an IIR are a simpler alternative; however, in these situations, estimators with an FIR, e.g. SG-filters, may also be suitable in some applications; for instance, when perfect phase-linearity in a causal realization is required. Designing low-pass and band-pass filters (FIR or IIR) in the frequency domain with derivative constraints at dc for vanishing moments in the time domain is an alternative way of deriving monomial-matched filters; however, these methods obfuscate the underlying statistical framework of linear least-squares regression upon which SG filters are based.

The taper of the impulse-response tails and the point at which they are truncated, are critical aspects of FIR filter design. Side-lobes cannot be eliminated in FIR filters; however, their structure may be altered by optimally shaping the impulse response. For a fixed polynomial degree, the side-lobes (and variance) of a conventional SG- filter are lowered by increasing the filter order; however, this also decreases bandwidth (and increases bias). Sequentially-correlated (i.e. 'colored') noise models are used here to set the bandwidth and side-lobe structure of SG-filters.

Narrow-band Gauss-Markov noise models place filter nulls that cancel known sources of interference; however, a wide-band high-frequency model is a more robust alternative in uncertain noise environments. The wide-band model is an alternative way of deriving band-limited SG-filters; narrow-band models have not previously been used in SG-filter designs. In both cases, the resulting linear-phase FIR filters may be used to perform smoothing and differentiation operations, or to generate feature-vectors for transient description and detection, in malign colored-noise environments.

`

`